\documentclass[12pt,preprint]{aastex}

\slugcomment{\today}

\begin{document}

\title{Spitzer-IRAC Search for Companions to Nearby, Young M Dwarfs}

\author{Peter R.\ Allen}
\affil{Pennsylvania State University, Dept.\ of Astronomy and Astrophysics, 525 Davey Lab, University Park, PA 16802; pallen@astro.psu.edu}
\author{I.\ Neill Reid}
\affil{\small Space Telescope Science Institute, 3700 San Martin Drive, Baltimore, MD 21218; inr@stsci.edu}

\begin{abstract}

We present the results of a survey of nearby, young M stars for wide low-mass companions with IRAC on the Spitzer Space Telescope.  We observed 40 young M dwarfs within 20~pc of the Sun, selected through X-ray emission criteria.  A total of 10 candidate companions were found with IRAC colors consistent with T dwarfs.  Extensive ground-based NIR follow-up observations rejected all these candidates.  Two additional candidates were discovered via common proper motion measurements, one of which was rejected as a background object and the other is a {\it bona fide} companion to GJ 2060, a member of the AB Doradus moving group.

\end{abstract}

\keywords{stars: low-mass, brown dwarfs; (stars:) binaries: visual; stars: formation}

\section {Introduction}

The past decade has seen an extraordinary increase in our understanding of the lower reaches of the main sequence. Stimulated largely by a new generation of near-infrared sky surveys, a wealth of low luminosity dwarfs has been uncovered, requiring the addition of two new spectral classes to the standard canon: L, spanning an approximate temperature range of $\sim2000$~K to 1300~K \citep{k99}, and T, methane dwarfs, reaching temperatures as low as $\sim700$~K \citep{burg02,burg06,geb02}. Many L dwarfs and all T dwarfs are substellar-mass brown dwarfs.

Current efforts focus on extending coverage to even lower temperatures, where spectral changes are expected to require yet another spectral class, provisionally designated Y \citep{k99}. Finding isolated examples of such objects in the general field is hampered by their low luminosities (M$_{bol}> 19$, M$_J > 17$) and correspondingly low surface densities in even deep near-infrared surveys, such as UKIDDS \citep{dye}. Even the optimistic assumption of equal numbers of brown dwarfs and stars (a mass function, $\Psi(M) \propto M^{-1}$) predicts only $\sim4000$ brown dwarfs within 25 parsecs, or a surface density of $\sim1$ per 10 sq. degrees \citep{inr04,kc07}.  At least half of those dwarfs are expected to have T$<$300~K, and absolute magnitudes M$_J>19$ and M$_{4.5{\mu}m}>18$ according to synthetic mass function estimates \citep{burg04,a05}, taxing ground and space-based surveys to their limit.

Field surveys are not the only option available to searches for low mass dwarfs. Indeed, the most successful method of extending the main sequence has been searching for low luminosity companions of stars that are already known to be close neighbors of the Sun. This approach led to the discovery of the first ultracool dwarf (Gl752B, or VB10, spectral type M8), as the companion of an M3 dwarf \citep{vb}; the first known L dwarf (GD 165B), a close companion of a DA white dwarf \citep{bz88}; and the first T dwarf (Gl 229B), the resolved companion of an M0.5 dwarf at a distance of 6 parsecs from the Sun \citep{naka}.  Y dwarfs are likely to be the next step.

Companion searches offer significant advantages in determining the fundamental properties of low mass dwarfs. Even if direct trigonometric parallaxes are unavailable, reliable distance estimates, and hence luminosities, can be derived using the photometric and/or spectroscopic properties of the main-sequence primary. Moreover, detailed observations of the primary can provide crucial information on the metallicity, kinematics and, particularly, the age of the system, allowing an estimate of the mass of a brown dwarf companion. Finally, an inventory of the frequency and mass ratio distribution of close, low-mass companions is vital to assessing the likely frequency of planetary systems. 

Optimizing detection probabilities requires careful consideration of two factors: the luminosity contrast ratio between components; and the likely angular separation of binary systems. Thus, while low-luminosity ultracool dwarfs are closest in luminosity to brown dwarf companions, only a handful of the known binaries have separations exceeding 15 AU \citep{me06ppv}. On the other hand, luminous G and K dwarfs, overwhelm brown dwarf companions at moderate and small separations.

We have chosen to focus on Spitzer observations of nearby, young, mid-type M dwarfs. Those stars provide a better contrast ratio for companion detection than higher-mass GK stars at moderate orbital separation for low-mass companions and a sufficiently high space density to provide statistically significant numbers in the immediate Solar Neighborhood. Spitzer 4.5~$\mu$m observations offer the potential of detecting companions $\sim7.5$ magnitudes fainter than the primary at separations exceeding $\sim15$ arcseconds. Brown dwarfs cool and fade on relatively rapid timescales, so targeting young systems maximizes the probability of detecting very low-mass companions. Our observations have not resulted in the detection of any new ultracool companions; however, we can set upper limits to the frequency of wide, low-mass companions to stars in this mass range.

The paper is organized as follows: Section 2 describes target selection, outlining how we use X-ray data to identify relatively young M dwarfs near the Sun; Section 3 discusses our Spitzer-IRAC data reduction procedures and candidate selection criteria; Section 4 outlines our follow-up data; Section 5 discusses our results; and Section 6 summarizes our conclusions.

\section {Target selection}

X-ray activity is a powerful indicator of stellar youth. Observations of open clusters show a consistent and relatively rapid decline in emission strength with increasing age at a given spectral type. This is illustrated in Figure 1, which plots X-ray data for members of TW Hydrae association ($\tau \sim 10$ Myrs; \citet{inr03}), the Pleiades ($\tau \sim 120$ Myrs; \citet{st94,m99}) and the Hyades ($\tau \sim 450$ Myrs; \citet{st81,rhm}), together with data for local field dwarfs \citep{hu99}. 

We are currently compiling a census of K and M dwarfs within 20 parsecs of the Sun, concentrating on late-type dwarfs within the 48\% of the sky covered by the 2MASS second release (\citet{inr04,rca07}, and references within). So far, we have identified over 950 stellar systems that are likely to lie within our distance limits, including 320 additions to the nearby star catalog.  Most of the latter stars are drawn from the Luyten Two-Tenths Catalog \citet{nltt}. We have identified active stars by cross-referencing this dataset against the ROSAT bright source catalogue \citet{v99}. Twenty percent of the stars in this sample have X-ray counterparts.

While almost all the stars in the current sample have visual magnitude measurements, most lack reliable optical colors; however, all, by definition, have near-infrared data. We have therefore used the X-ray/J-band flux ratio as an activity indicator (Figure 2).  Our goal is to target stars that are likely to be younger than a few hundred Myrs, thereby optimizing the chances of detecting very low-mass companions.  Using data for Pleiades stars as a guide, we have selected stars with (V-J)$>$3.0 and flux ratios
\begin{displaymath}
\log {f_X \over f_J} \quad > \quad 0.33 \times (V-J) - 3.8
\end{displaymath}
Seventy-nine stars meet these criteria. Thirty-six of these active stars (40\%) are known to be in binary systems. Activity is often enhanced by interactions and spin-orbit coupling in close binaries, and we also wish to avoid secondary stars with earlier-type primaries, so we have excluded those stars from the Spitzer observations. 

Of the remaining 43 apparently single, active M dwarfs (spectral types M0 to M6), two (AD Leo and AU Mic) are included in Spitzer GTO programs. The remaining 41 stars were targeted for observations in the present program (see Table \ref{tab:targs}). 

Given the dispersion in activity at a given age, we do not expect all of these stars to be as young as the Pleiades. However, these stars are likely to include the youngest constituents of the immediate Solar Neighborhood and, as a consequence, offer the best prospects of detecting extremely low-mass brown dwarf companions.

\section{Spitzer-IRAC Data}

\subsection{Observations}

Our Spitzer-IRAC observations of 40 out of 41 targets were obtained throughout Cycle-1 (see Table \ref{tab:targs}).  One target, Wo 9520, was not observed because, for an unknown reason, that observation was replaced by a repeat observation of a previously observed target.  We adopted short exposure times of 12~s to minimize persistence problems, since the primaries are all bright ($7<K<9$).  The 40 targets were observed in a combination of full-frame and high dynamic range (HDR) mode.  HDR mode was used for the 10 brightest targets.  This mode takes a short 0.6~s exposure followed by the long 12~s exposure at the same dither position.  In this way, we can obtain accurate photometry for bright sources that are saturated in the longer exposures.  Those targets observed in HDR mode are indicated in Table 1.  All other sources were observed in full frame mode with 12~s exposures at each pointing.  We used a 9-point random dither pattern, as prescribed in the Observer's Manual, to remove detector defects and improve our angular resolution.  The small-scale dither pattern was used to optimize sub-pixel sampling.

\subsection{Data Reduction}

The Basic Calibrated Data were downloaded from the Spitzer archive via Leopard.  We then used the Spitzer Science Center provided data reduction software MOPEX/APEX, to carry out our analysis.  MOPEX was used to mask bad pixels and cosmic rays, smooth over electronic artifacts such as column pulldown and muxbleed, shift and add the dithered images, and perform aperture photometry. The HDR mode data was reduced separately from the full range mode data.  Sources that were saturated in the long exposures were identified and photometry taken from the unsaturated shorter exposures, as available.  

Final band-merged photometry lists for each field were generated by requiring identification of each source in at least the 3.6~$\mu$m and 4.5~$\mu$m channels.  This eliminates many artifacts automatically.  We also required sources to have magnitude uncertainties better than 0.15~mag as an additional quality check.  The final factor taken into account is distance from the primary star; we exclude sources within a 15\arcsec~radius of the bright primary stars due to the likelihood of contamination by the primary PSF.  As Figure \ref{fig:psf} shows, the inner boundary of this search region lies on the wings of the Spitzer PSF at 3.6~$\mu$m.  As a result, we effectively have a uniform sensitivity over the full area used for the companion search.  The photometry lists for sources at separations greater than 15\arcsec~were then used to select candidate companions.

\subsection{Companion Candidate Selection}

The candidate selection was a multi-step process.  We used the calibrated aperture photometry derived by MOPEX/APEX, described above, and FORTRAN scripts to parse the data.  This resulted in 14621 sources.  Next, all the sources detected at both 3.6~$\mu$m and 4.5~$\mu$m were compared to the 2MASS database \citep{2mass} via a Gator query within 5 arcseconds of each IRAC source.  This search yielded 2513 2MASS sources.  

The primary purpose of this project is to search for very cool brown dwarf companions to known nearby M dwarfs.  However, with the addition of the NIR 2MASS photometry it becomes possible to distinguish spectral type M and L companions from even hotter stars (FGK).  M and L dwarfs have neutral IRAC colors, but have red 2MASS-IRAC colors which allows us to identify them photometrically.  In addition, matching the IRAC data against 2MASS allows us to determine whether candidate photometric companions display common proper motion (CPM) with their putative primary stars.

All sources detected in both our IRAC data and 2MASS were matched against color and magnitude criteria consistent with a M, L, or T dwarf.  Those criteria are based on Spitzer-IRAC photometry of known MLT dwarfs described in \citet{pat}.  One of the cleanest selection techniques is to plot the $M_{Ks}$ versus $K_{s}-4.5~\mu$m color-magnitude diagram.  To construct this diagram, the sources in each field were assumed to lie at the distance of the targeted M dwarf.  Distance estimates are predominantly derived from photometric and spectroscopic parallaxes, with only 8 sources having Hipparcos parallaxes, see Table \ref{tab:targs}.  The results are displayed in Figure \ref{fig:akvkm2}.  Of the 2513 sources with 2MASS counterparts only one has colors and magnitudes consistent with a MLT dwarf (see Table \ref{tab:tmc}).  However, this object does not display common proper motion (CPM) with its putative primary, Gl 173-29, and, given its 2MASS colors, is likely a white dwarf.

We have also searched for CPM companions that lie in otherwise crowded areas of color space.  In our analysis we found two candidate companions, and also independently reacquired the wide M3 companion to G~274-24, as reported in \citet{jao}.  The first candidate companion is in the field of LP 764-40, has neutral 2MASS-IRAC colors (Table \ref{tab:tmc}), and is visible on scans of the second Palomar Sky Survey plate material (DPOSS).  A low-resolution NIR spectrum was obtained for us, by Michael Liu and Katelyn Allers, using Spex on NASA's IRTF.  This spectrum is consistent with a M4 dwarf, implying a distance of several hundred parsecs.  This compares with a distance of 18.6~pc to the putative primary, and indicates that this M4 dwarf cannot be physically associated with LP 764-40.  The second CPM candidate companion lies in the field of GJ 2060.  This object, as will be discussed further in Section \ref{sec:gj2060}, is found to be a genuine companion.  Thus, all sources with 2MASS counterparts, save one, are ruled out as potential companions.

For sources with IRAC-only detections, it is not possible to distinguish companions warmer than T dwarfs, see Figure \ref{fig:iraccrit}, so we focus our efforts here on T and cooler candidates.  However, we can eliminate sources with IRAC-only colors consistent with dwarfs earlier than T0 since those sources ought to have JHK magnitudes brighter than the 2MASS limits if they actually were late-type companions of the target star.  Thus, those sources are also eliminated (Figure \ref{fig:iraccrit}).  Again, our selection criteria are based on the IRAC photometry of known MLT dwarfs as described in \citet{pat}.  Our first cut was in $M_{4.5~\micron}$ versus $3.6~\micron - 4.5~\micron$ (Figure \ref{fig:iraccrit}).  These criteria are defined for all objects with $3.6~\micron - 4.5~\micron > -0.35$ 

\begin{eqnarray}
\nonumber M_{4.5~\micron} \leq 1.25{\times}[3.6~\micron - 4.5~\micron] + 13.0\\
M_{4.5~\micron} \geq 1.25{\times}[3.6~\micron - 4.5~\micron] + 9.0
\end{eqnarray}

A total of 45 candidates survived these cuts.  Two further color cuts were applied, again based on Patten et al.  For those sources detected at $5.8~\micron$ as well, we eliminated all sources with $4.5~\micron - 5.8~\micron$ colors greater than 0.6, see Figure \ref{fig:iraccrit2}.  Finally, for those sources also detected at $8.0~\micron$, we eliminated all those objects with $3.6~\micron - 4.5~\micron < 2$ and with $5.8~\micron - 8.0~\micron$ colors red-ward of the T dwarf sequence, see Figure \ref{fig:iraccrit3}.  The last step was to compare the remaining candidates with the DPOSS plates.  Since L and T dwarfs are very red, we would not expect a genuine low-mass dwarf to appear on the plates.  Thus, any candidate seen on the DPOSS plates was eliminated.  The final number of sources remaining after all of these cuts is only 10 sources, which are listed in Table \ref{tab:final}.

\section{Follow-up Observations}

\subsection{IRAC Only Candidates}

The most efficient means of establishing the nature of the IRAC-only candidates is to obtain NIR JHK photometry.  The latter observations were carried out thanks to generous donations of time on three different telescopes, as indicated in Table \ref{tab:final}.  One candidate was observed using the imaging mode of Spex on the IRTF in Hawaii in June 2005.  The majority of the remaining observations (7 out of 10) were performed using the PANIC instrument on the Magellan telescope in Chile in November 2005.  The two final candidates were observed using SWIRC on the MMT in Arizona also in November 2005.  All of these candidates were either resolved as galaxies (4/10) or were found to have NIR-IRAC colors that are inappropriate for T (or Y) dwarfs (6/10).  Those objects with wrong NIR-IRAC colors are all found to be too red, particularly in J-K.  Given that nearly half of the candidates were resolved into galaxies and the very red NIR-IRAC colors, these unresolved objects are likely galaxies as well.  Thus, we conclude that none of the IRAC only sources are confirmed as a cool brown dwarf companion of a nearby young M dwarf.

\subsection{GJ 2060}
\label{sec:gj2060}
As mentioned in Section 3.3, one object in the field of GJ 2060 displays CPM with the primary.  GJ 2060 (2MASS J07285137-3014490) is a M1 dwarf at a distance of 15.6 parsecs, which was recently resolved as a close binary {\sl via} ground-based AO images \citep{daem07}. GJ 2060 has also been shown to be a likely member of the nearby $\sim$50~Myr old stellar association AB Dor \citep{abdor}.  

The primary reached the hard saturation limit in both the 3.6$\mu$m and 4.5$\mu$m channels (this field was not observed in HDR mode) and is mildly saturated in the remaining channels.  The companion is also nearing the hard saturation limit in all four channels of our IRAC observations.  This renders the photometry for the primary as well as that of the secondary uncertain, see Table 1.  

However, accurate astrometry can be obtained because short 0.6s exposures are taken at the beginning of IRAC dither sequences.  Due to the nature of how the data is taken, these short exposures are only available in the 4.5$\mu$m and 8.0$\mu$m channels.  Matching the unsaturated IRAC astrometry against 2MASS data leads to derived motions that are consistent between the primary and secondary given the 10-20 mas/yr uncertainties, see Table \ref{tab:gj}.  The 2MASS NIR colors (Table \ref{tab:gj}) and the location on the M$_J$ v J-K color-magnitude diagram indicate that this CPM companion is likely a mid M dwarf, which we identify as GJ 2060C.  

To test this hypothesis, a NIR spectrum of the candidate was obtained with the Spex instrument on the IRTF in November 2006.  The spectrum was reduced using the Spextool software package \citep{spextool} and was found to be that of a mid-M dwarf.  We overlaid several standard spectra from the IRTF spectral library \citep{cush05} and determined that the best fit spectral type is M5V (Figure \ref{fig:spec}). As a companion, the angular separation of $67\farcs2$ corresponds to a linear projected separation of 1050~AU at the Hipparcos distance of 15.6~pc.  However, the new companion is over-luminous for the distance of the system and a M5 spectral type.  A M5 dwarf has a nominal absolute magnitude of M$_J$ = 9 - 9.5 \citep{dahn}, whereas GJ 2060C has $M_J\sim$8 for $d\sim15.6$ parsecs, implying that it is over-luminous by a factor of 2.5-4.  The most likely explanation is that the companion is itself a binary system.  

Thus, we conclude that GJ 2060 is probably a quadruple system, composed of two widely separated equal-mass binaries. The combination of a good age estimate, the nearby distance, and the ability to obtain dynamical masses makes this an extremely interesting system for further study.  Mass and age are two of the most important and elusive stellar parameters to measure.

\section{Discussion}

This survey is complementary to the one carried out by \citet{daem07}.  The source sample is the same, but the data considered here are sensitive to companions at much wider separations.  We excluded the area of each image around the target star to a radius of 15\arcsec (Figure \ref{fig:psf}).  At this radius we have uniform sensitivity from the bright target stars out to the edge of the field, or about 2.5 arcminutes.  This corresponds to separations of $\sim$200~AU to $\sim$2200~AU at our typical target distance of 15~pc.  This separation range is also complimentary with the $\sim$300~AU limit of the \citet{daem07} work.  We update the overall results from the ground-based survey with the discovery of the quadruple nature of the GJ 2060 system.  Thus the total number of multiple systems of the 40 observed star sample becomes: 11 binaries, 1 triple, and 1 quadruple, where the quadruple is GJ 2060.

Most of our sample yielded null results (38/40), thus it is important to characterize our sensitivity limits.  The ultimate limiting factor in our selection of candidate companions is the $3.6~\micron$ magnitude.  This is because the final selection criteria used was the $3.6~\micron-4.5~\micron$ color.  While IRAC's $4.5~{\mu}m$ channel is less sensitive than the $3.6~\micron$ channel in terms of raw numbers ($15~{\mu}$Jy versus $12~{\mu}$Jy for $3~\sigma$ detections), we also need to consider the SED of a T dwarf.  They are considerably fainter in $3.6~\micron$ than in $4.5~\micron$, with colors greater than $0.6-0.7$ for mid to late T dwarfs, as well as for projected cooler spectral types \citep{bur03}.  However, the reddest color that we can detect near our survey limits corresponds to ${\sim}0.25$ mag.  Thus, since we require detections in both the $3.6~\micron$ and $4.5~\micron$ channels the actual limiting magnitude is in the $3.6~\micron$ channel.  

The limiting magnitudes in each field were determined automatically using the same photometry used to select candidates.  The maximum magnitude detected in each field was determined by combing through the combined sources lists and looking for sources with magnitude errors less than 0.15, which were used in the selection process.  The derived magnitude limits for each field are listed in Table \ref{tab:targs}.  The median values are around $m_{3.6} \sim 17.5$ and $m_{4.5} \sim 17$.  

The more interesting limiting quantity is the mass.  However, in order to establish a mass limit we must use evolutionary models.  These models are an essential ingredient in the study of substellar objects because brown dwarfs do not maintain stable hydrogen fusion.  Thus, their luminosity is heavily age dependent and, given the lack of empirical measurements, the models are required to obtain mass estimates.  We apply the same approach as in \citet{a05}, using the evolutionary models of \citet{bur} and the bolometric corrections from \citet{golim}.  The bolometric corrections are required to transform the bolometric luminosity provided by the evolutionary models to an observed band pass magnitude.  We used the \citet{golim} L' and M band bolometric corrections to stand in for IRAC channels 3.6~\micron and 4.5~\micron because there are no published corrections for IRAC channels.

We determined the mass that corresponds to the absolute limiting magnitude of each field for three different ages: 120 Myr (Pleiades-like); 500 Myr (Hyades-like); and 5000 Myr (field age).  The first two ages were chosen because our sample is drawn from stars that were selected to have ages comparable to, or less than, that of the Hyades and are likely greater than the Pleiades.  We also consider an average age for the general Solar Neighborhood for comparison.  The resulting histograms of limiting masses are displayed in Figure \ref{fig:masslim}.  Not surprisingly, the limiting mass is lower with a younger age; the younger a brown dwarf is the brighter it shines.  The median limiting mass at Pleiades ages is $\sim2$ Jupiter masses, $\sim6$ Jupiter masses at Hyades ages, and $\sim18$ for the field.  Conservatively, we will take the 500 Myr Hyades age as representative of the sample.  Thus, our survey is complete for companions with masses as low as 6 Jupiter masses for separations from 200~AU to 2200~AU.

It is also interesting to examine the subset of the sample that were resolved into close binaries in \citet{daem07}.  Of the 12 binary systems reported it turns out, when combined with the present results, that 2 of those systems have tertiary components.  This is a wide tertiary fraction of $\sim15\%$, but this number should be taken with extreme caution for we are dealing with very small number statistics.  

These results are interesting in the light of recent numerical simulations of star formation.  These models are of the collapse of small, isolated, turbulent clouds and the subsequent dynamical evolution of stars in multiple systems.  In the simulations of \citet{dd04}, they predict that if a low-mass dwarf is found to be in a stable, wide ($>$10~AU) orbit its primary is frequently ($\sim75\%$) a tight binary.  A similar result is found in the dynamical simulations of \citet{ster}.  In addition, the \citet{dd04} simulations find that around 10\% of their tight multiple systems survive with wide ultracool dwarf companions by the end of their simulations, 10.5~Myr.  The surviving fraction of 10\% is highly suggestive in the light of our wide fraction of 15\%.  If true then the wide binary fraction could be used as a test of different star formation simulations.  However, much more observational work must be done to determine the nature of this trend.  This work is beginning in \citet{tok06}, who have examined 165 spectroscopic binaries and found an additional tertiary frequency of $63\%\pm5\%$.  They also determined that this frequency is a strong function of the period of the tight binary pair.  Accurate measurements of the companion distribution of these complex systems will be required to fully answer questions about star formation processes.

\section{Summary}

We have conducted a survey of 40 nearby, young M dwarfs using IRAC on the Spitzer Space Telescope to find wide, ultracool companions.  The target stars were chosen to have X-ray activity levels that exceed Pleiades stars of similar spectral type, and are therefore likely to have ages of a few hundred Myrs.  Our observations are capable of detecting late-M, L, or T dwarfs at separations more than 15 arcseconds from the target stars, which lie at distances of 9 to 22 parsecs.  No ultracool companions were detected in our sample, but we discovered a new quadruple system in GJ 2060.  The known wide triple, G 274-24, we also reconfirmed in our CPM analysis.  Our sensitivity limits indicate that we could have detected most if not all companions down to masses of 6 Jupiter masses.  There is also some limited evidence in support of dynamical simulations of \citet{ster} and \citet{dd04}.  However, this conclusion is based on very small number statistics and will require substantial additional observations to confirm.

{\noindent {\it Acknowledgments} This work is based in part on observations made with the Spitzer Space Telescope, which is operated by the Jet Propulsion Laboratory, California Institute of Technology under a contract with NASA.  Support for this work was provided by NASA through an award issued by JPL/Caltech.  PRA acknowledges support from grant NAG5-11627 from the NASA Long-Tern Space Astrophysics program made to Kevin Luhman.}

\noindent This publication makes use of data products from the Two Micron All Sky Survey, which is a joint project of the University of Massachusetts and the Infrared Processing and Analysis Center/California Institute of Technology, funded by the National Aeronautics and Space Administration and the National Science Foundation.  This research has made use of the SIMBAD database, operated at CDS, Strasbourg, France.  

\noindent The authors would like to thank warmly several people for obtaining critical follow-up observations: Michael Liu and Katelyn Allers for Spex imaging and spectroscopy using NASA's IRTF; Brian Patten for SWIRC imaging using MMT; Adam Burgasser for PANIC imaging using Magellan; and Davy Kirkpatrick and Dagny Looper for the IRTF Spex spectrum of GJ 2060C.

\clearpage

\begin{figure}
\plotone{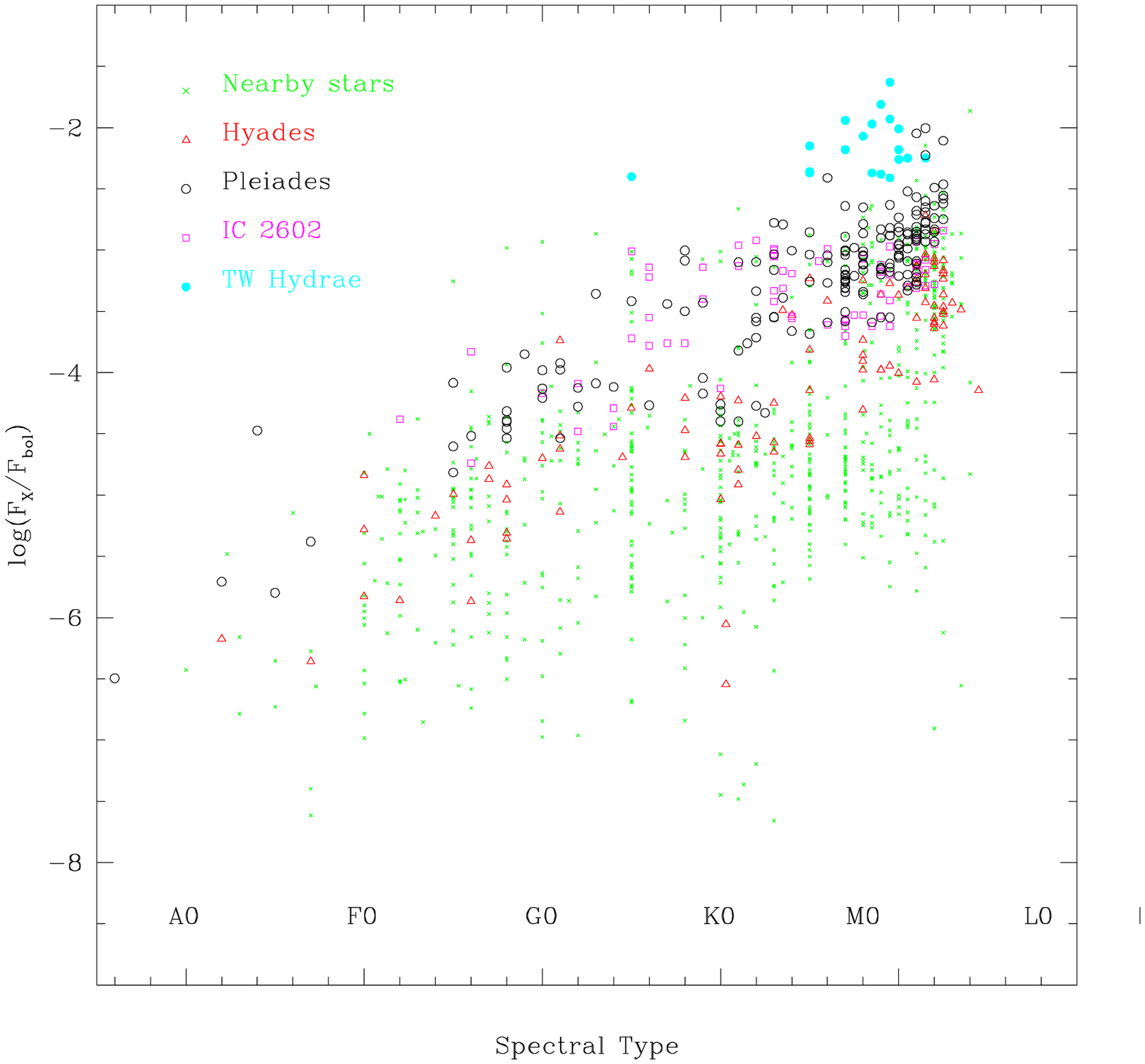}
\caption{Coronal activity as a function of spectral type.  The crosses mark data for field stars in the immediate Solar Neighborhood (from \citet{hu99}); open triangles are data for the Hyades cluster (500~Myr) \citep{rhm,st81}; open circles are data for the Pleiades (120~Myr) \citep{st94,m99}; open squares are data for IC~2602 (50~Myr) \citep{rand95}; and solid circles mark data for members of the TW Hydrae Association (10~Myr) \citep{inr03}.}
\end{figure}

\clearpage

\begin{figure}
\plotone{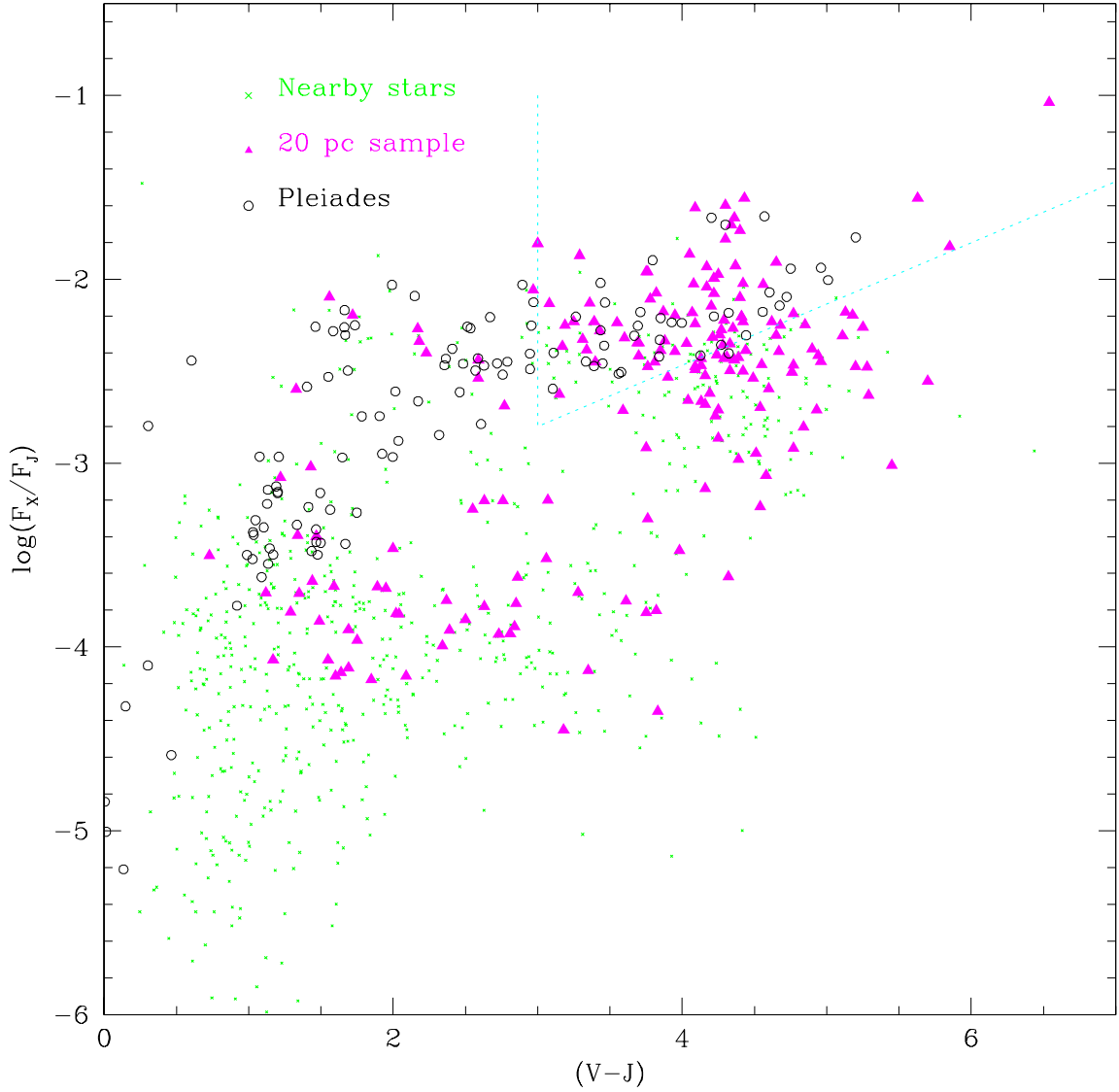}
\caption{X-ray selection of young nearby stars: we have computed the
X-ray/J-band flux ratio for the 196 stars in our 20-parsec sample that are included in the ROSAT bright source catalogue, and plot those data against (V-J) color (solid triangles). Similar data are shown for nearby stars from the Huensch et al catalog (crosses) and for Pleiades stars from Micela et al's HRI survey (open circles). The dotted lines outline the selection criteria that we have used to select the 20-parsec M dwarfs most likely to have Pleiades-like ages. }
\end{figure}

\begin{figure}
\centering
\includegraphics[scale=0.9,angle=-90]{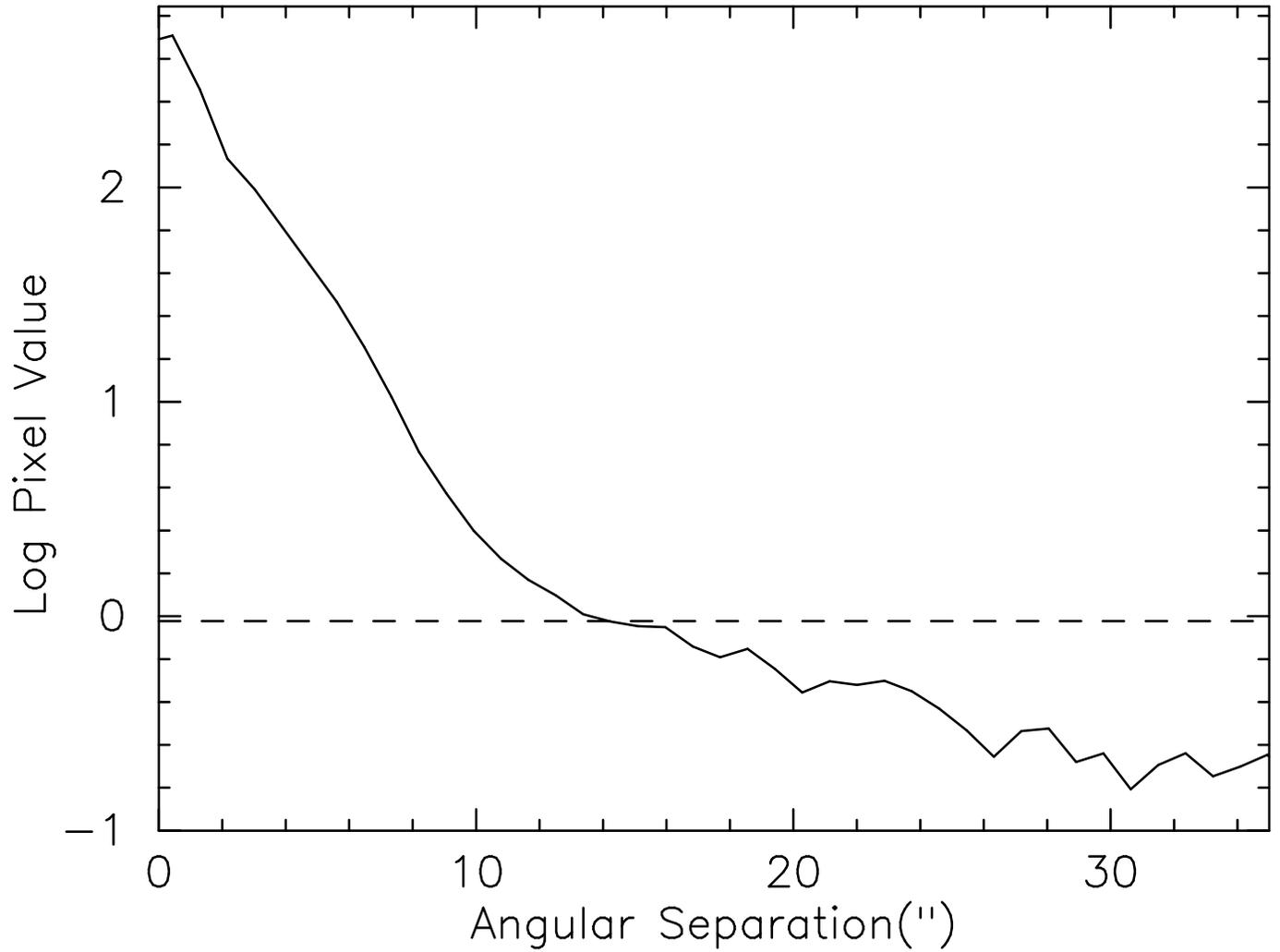}
\caption{Example PSF derived from the 3.6~\micron~image of LP~776-25 (solid line) with the 5$\sigma$ detection threshold (dashed line).  The logarithm of the pixel value has been taken to more easily see the point at which the detection threshold is crossed.}
\label{fig:psf}
\end{figure}

\begin{figure}
\centering
\includegraphics[scale=0.9,angle=-90]{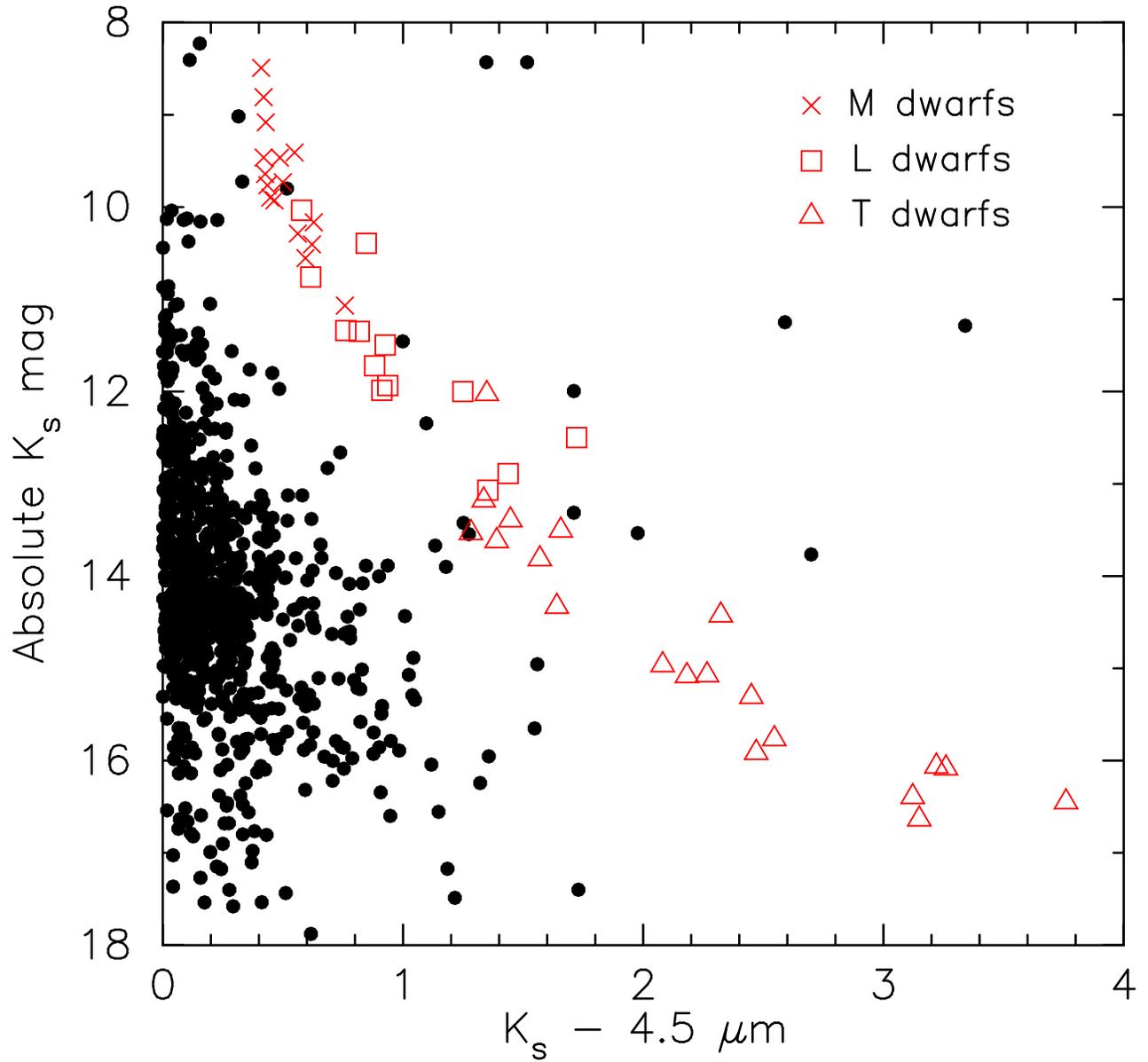}
\caption{$M_K$ versus $K-4.5~\micron$ for all IRAC sources detected in 2MASS (black circles) and known MLT dwarfs observed by the IRAC GTO team in Patten et al.\ (red shapes).  Note how well the entire MLT dwarf regime is separated from the bulk of sources in this color-magnitude diagram.}
\label{fig:akvkm2}
\end{figure}

\begin{figure}
\centering
\includegraphics[scale=0.9,angle=-90]{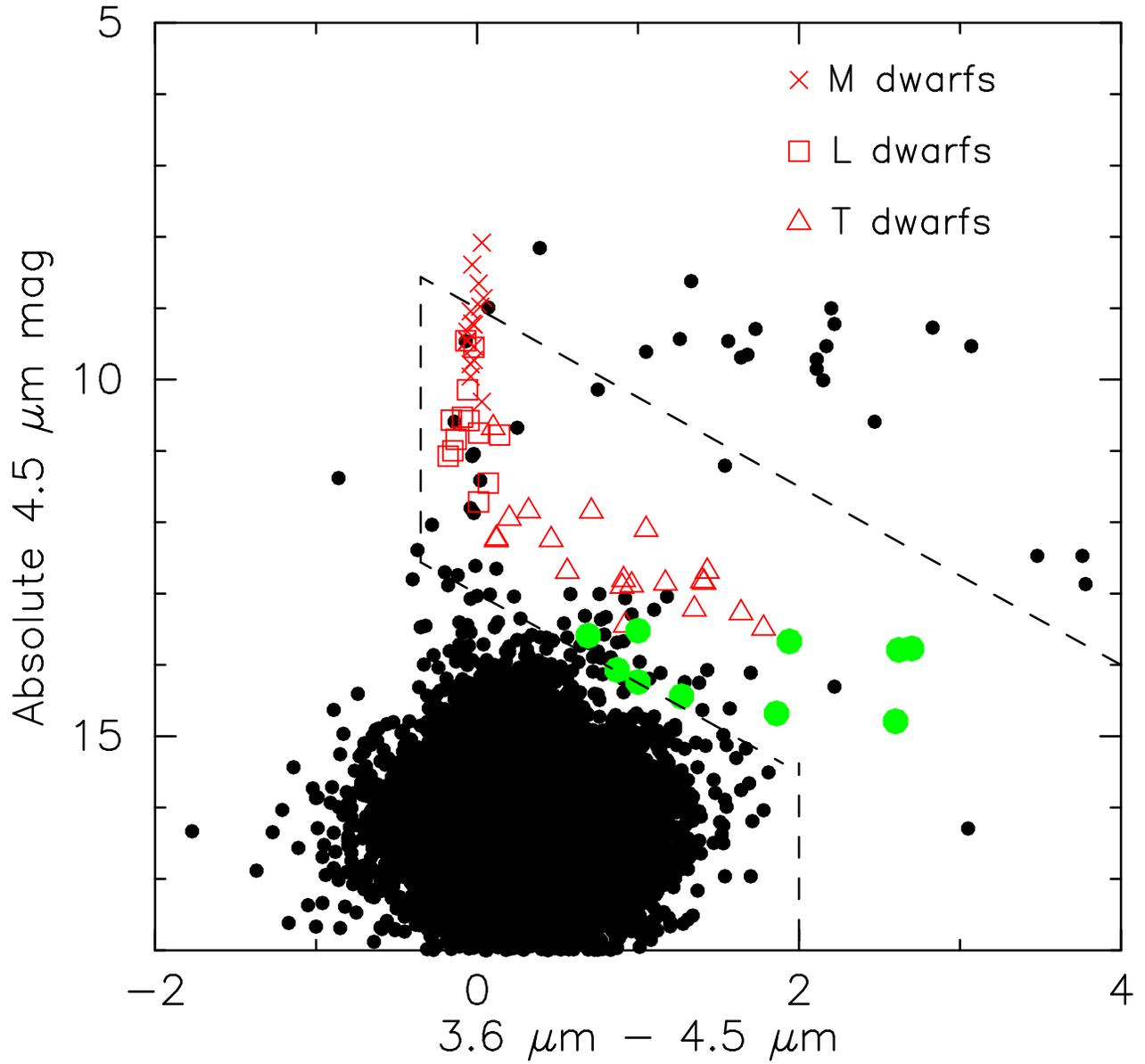}
\caption{$M_{4.5~\micron}$ versus $3.6~\micron - 4.5~\micron$ color-magnitude diagram of all IRAC sources detected in $3.6~\micron$ and $4.5~\micron$ and not detected in the 2MASS database (black circles), known MLT dwarfs from Patten et al.\ (red shapes), and the initial selection criteria used (dashed lines).  The final 10 IRAC-only candidates are also highlighted (large green circles).}
\label{fig:iraccrit}
\end{figure}

\begin{figure}
\centering
\includegraphics[scale=0.9,angle=-90]{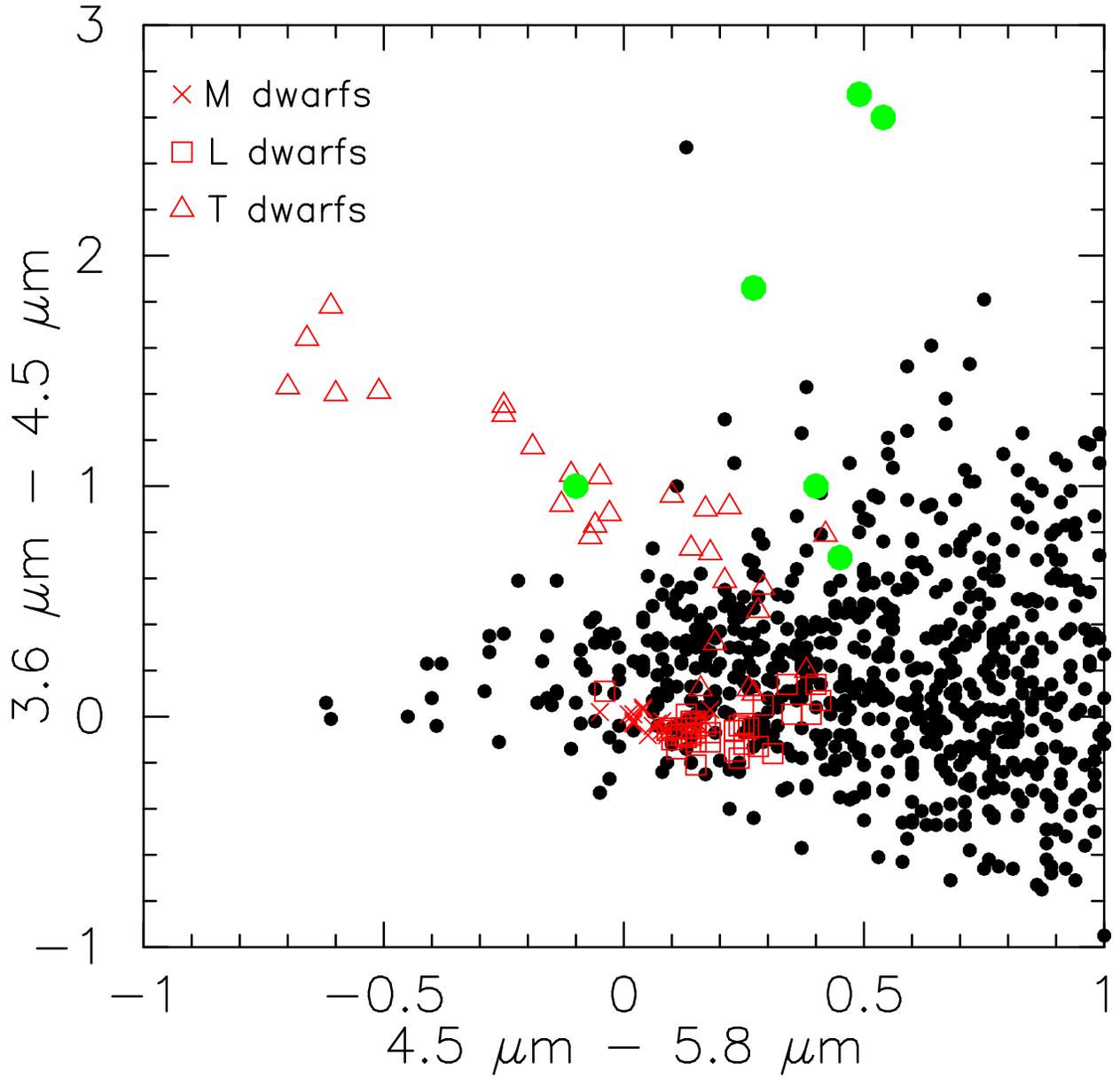}
\caption{$4.5~\micron - 5.8~\micron$ versus $3.6~\micron - 4.5~\micron$ color-color diagram of all IRAC sources detected in $3.6~\micron$ and $4.5~\micron$ and not detected in the 2MASS database.  All symbols are the same as in Figure \ref{fig:iraccrit}.}
\label{fig:iraccrit2}
\end{figure}

\begin{figure}
\centering
\includegraphics[scale=0.9,angle=-90]{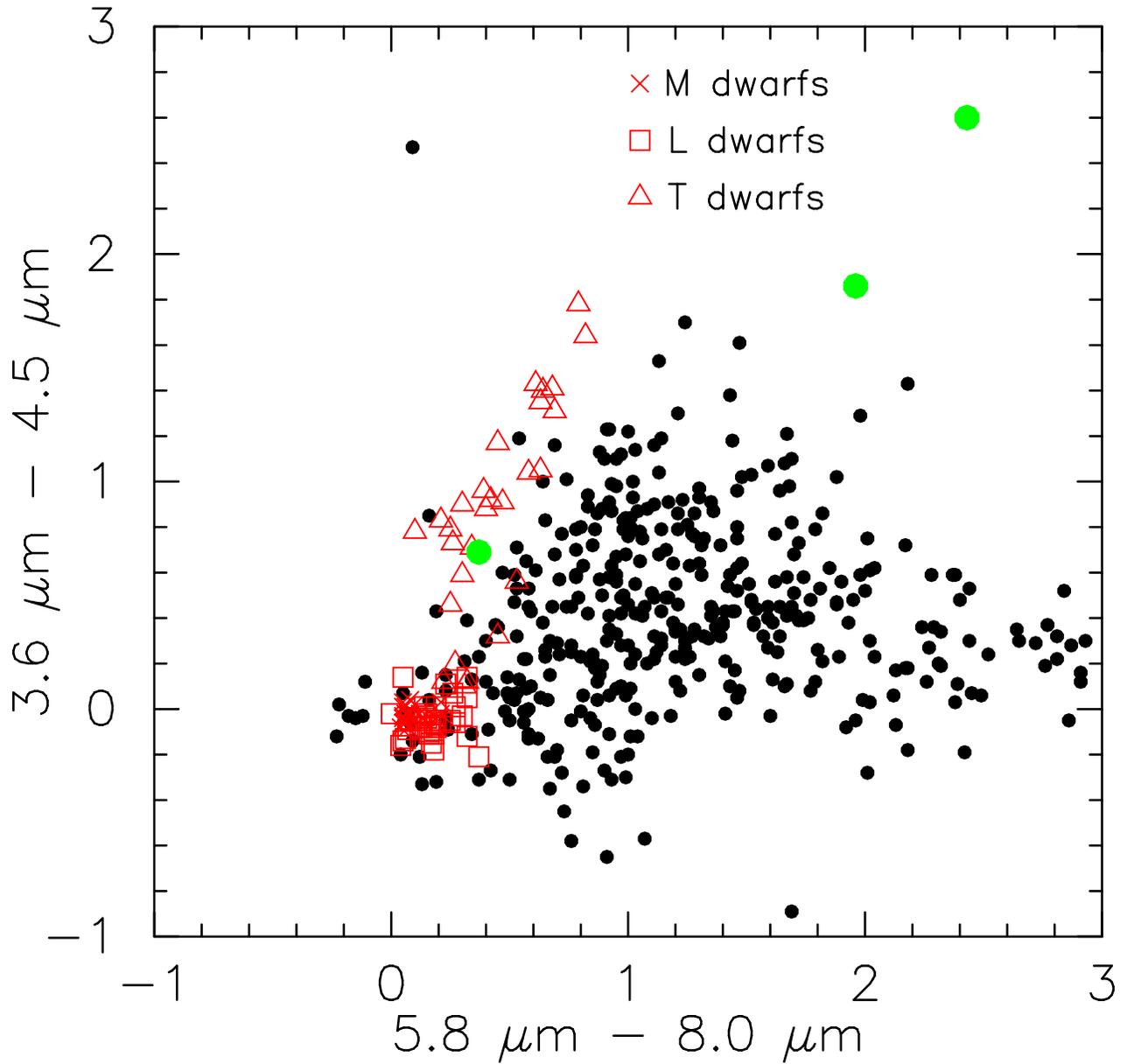}
\caption{$5.8~\micron - 8.0~\micron$ versus $3.6~\micron - 4.5~\micron$ color-color diagram of all IRAC sources detected in $3.6~\micron$ and $4.5~\micron$ and not detected in the 2MASS database.  All symbols are the same as in Figure \ref{fig:iraccrit}.}
\label{fig:iraccrit3}
\end{figure}

\begin{figure}
\centering
\includegraphics[scale=0.9,angle=-90]{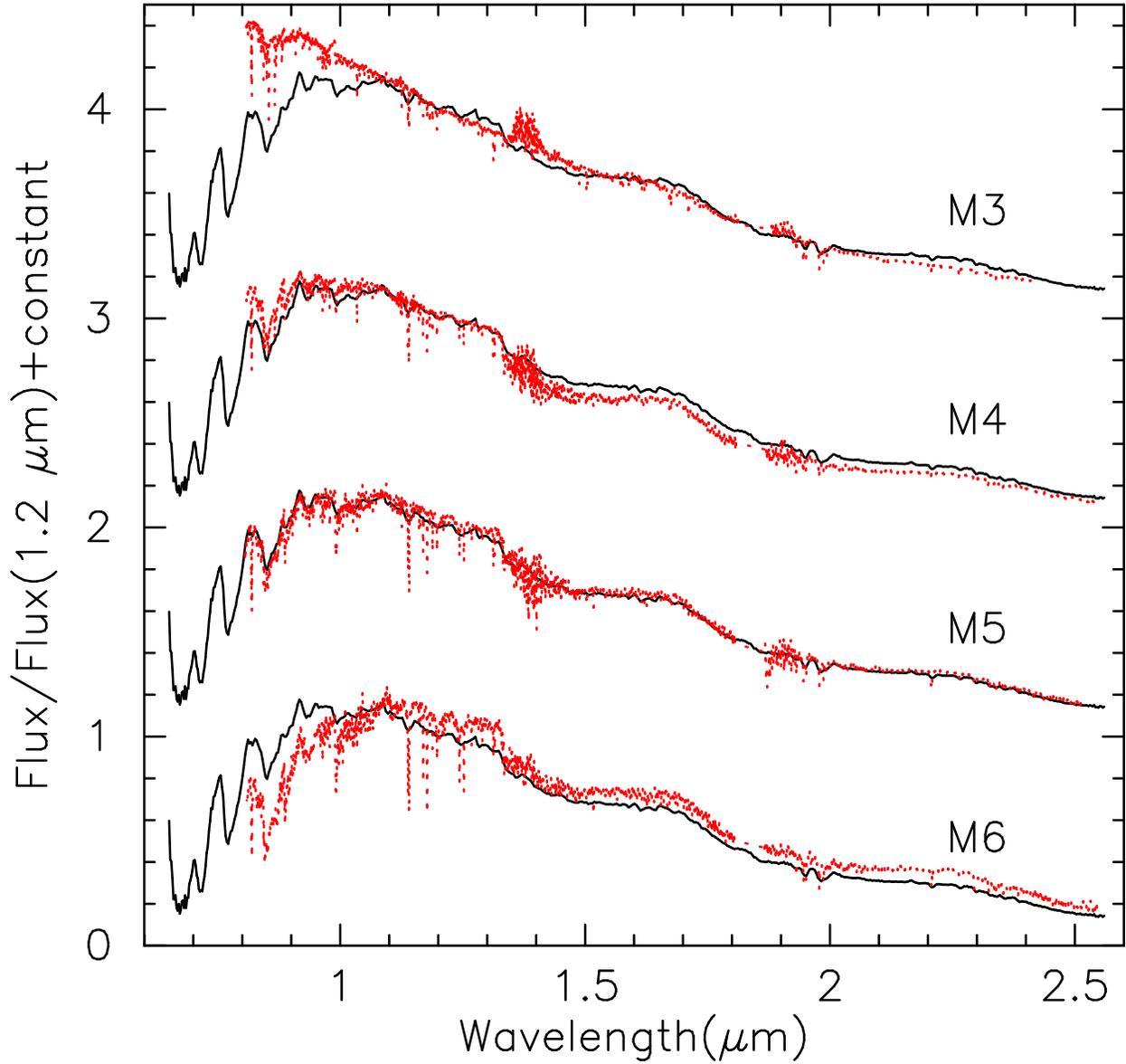}
\caption{Low-resolution NIR Spex spectrum of GJ 2060C (black solid lines) and M3-M6 standard spectra (red dashed lines) for comparison.  Each standard spectrum and that of the candidate is normalized to one at 1.2\micron.  The best fit is the M5 standard.}
\label{fig:spec}
\end{figure}

\begin{figure}
\centering
\includegraphics[scale=0.9,angle=-90]{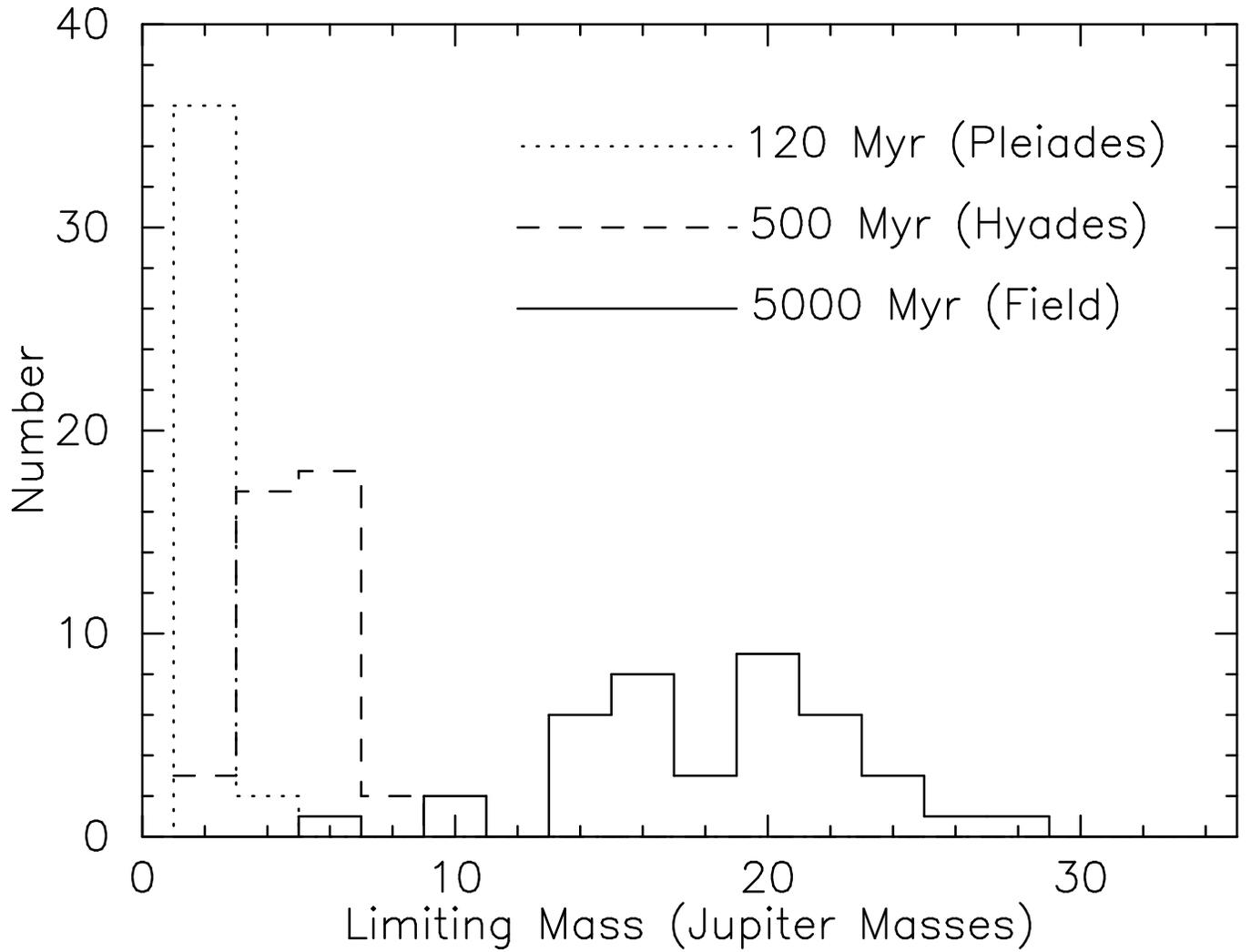}
\caption{Histograms of the mass limits of each target field for three ages: Pleiades like (120 Myr, dotted line); Hyades-like (500 Myr, dashed line); and field age (5000 Myr, solid line).  }
\label{fig:masslim}
\end{figure}

\begin{deluxetable}{lccccrrccccc}
\tabletypesize{\footnotesize}
\rotate
\tablewidth{0pt}
\tablecaption{Targets}
\tablehead{
\colhead{Name} & \colhead{$m_{3.6}$} & \colhead{$m_{4.5}$} & \colhead{$m_{5.8}$} & \colhead{$m_{8.0}$} & \colhead{RA} & \colhead{DEC} & \colhead{$m_{3.6}$ limit} & \colhead{$m_{4.5}$ limit} & \colhead{$m-M$} & \colhead{AOR ID} & \colhead{Notes} \\
\colhead{} & \colhead{(mag)} & \colhead{(mag)} & \colhead{(mag)} & \colhead{(mag)} & \colhead{(hms)} & \colhead{(dms)} & \colhead{(mag)} & \colhead{(mag)} & \colhead{(mag)} & \colhead{} & \colhead{} }

\startdata

LP 348-40  & 8.58  & 8.16  & 7.65  & 7.61  & 00 11 53.0 & $+$22 59 04 & 18.0    & 17.0     & $ 1.02\pm0.37$  & 10698752 & \\
G 172-001  & 8.49  & 8.10  & 7.63  & 7.61  & 00 28 53.9 & $+$50 22 32 & 17.9    & 17.2     & $ 0.56\pm0.25$  & 10693376 & 1 \\
G 274-24   & 8.49  & 8.19  & 7.94  & 7.91  & 01 24 27.6 & $-$33 55 08 & 17.7    & 17.1     & $ 1.51\pm0.30$  & 10694144 & 1 \\
LP 828-89  & 8.49  & 7.47  & 6.95  & 6.89  & 01 53 11.3 & $-$21 05 43 & 17.6    & 16.6     & $ 1.53\pm0.22$  & 10701056 & \\
G 003-035  & 8.81  & 8.65  & 8.43  & 8.37  & 02 02 44.2 & $+$13 34 33 & 17.6    & 16.7     & $ 1.07\pm0.29$  & 10694400 & \\
G 173-039  & 8.58  & 7.62  & 7.26  & 7.21  & 02 08 53.5 & $+$49 26 56 & 18.1    & 17.3     & $ 0.68\pm0.23$  & 10693632 & \\
Gl 103     & 5.39  & 5.06  & 4.81  & 4.68  & 02 34 22.5 & $-$43 47 46 & 17.3    & 17.1     & $ 0.30\pm0.02$* & 10695680 & 2 \\
G 36-26    & 9.07  & 8.96  & 8.77  & 8.76  & 02 36 44.0 & $+$22 40 26 & 17.6    & 17.0     & $ 0.97\pm0.31$  & 10694656 & \\
HIP 17695  & 6.79  & 6.76  & 6.68  & 6.59  & 03 47 23.2 & $-$01 58 19 & 18.3    & 17.2     & $ 1.06\pm0.11$* & 10696960 & 2 \\
II Tau     & 8.90  & 8.70  & 8.61  & 8.56  & 03 49 43.2 & $+$24 19 04 & 17.9    & 17.3     & $ 1.55\pm0.40$  & 10697216 & \\
Steph 497  & 6.31  & 6.32  & 6.29  & 6.22  & 04 37 37.4 & $-$02 29 28 & 17.8    & 17.0     & $ 0.99\pm0.22$  & 10702080 & 2 \\
G 039-029  & 8.46  & 7.74  & 6.97  & 6.89  & 04 38 12.5 & $+$28 13 00 & 18.0    & 17.0     & $-0.15\pm0.25$  & 10694912 & 1,3 \\
LP 655-48  & 9.29  & 9.23  & 9.15  & 8.99  & 04 40 23.2 & $-$05 30 08 & 17.9    & 17.2     & $ 0.07\pm0.19$  & 10698240 & \\
LP 776-25  & 6.91  & 6.67  & 6.74  & 6.63  & 04 52 24.3 & $-$16 49 21 & 17.6    & 17.1     & $ 0.45\pm0.22$  & 10700800 & 2 \\
LP 717-36  & 8.51  & 7.83  & 7.29  & 7.23  & 05 25 41.5 & $-$09 09 12 & 17.4    & 17.1     & $ 0.72\pm0.22$  & 10699776 & 1 \\
G 108-4    & 8.71  & 8.43  & 8.31  & 8.21  & 06 29 50.2 & $-$02 47 45 & 18.0    & 17.4     & $ 1.54\pm0.23$  & 10692608 & \\
GJ 2060    &  sat  &  sat  & 5.62  & 5.73  & 07 28 51.4 & $-$30 14 48 & 18.0    & 17.5     & $ 0.96\pm0.09$* & 10695424 & 1,4 \\
LHS 5134   & 6.11  & 6.18  & 6.08  & 5.98  & 08 08 13.6 & $+$21 06 09 & 17.6    & 17.1     & $ 1.17\pm0.10$* & 10697984 & 2,5 \\
Gl 316.1   & 9.79  & 9.73  & 9.59  & 9.53  & 08 40 29.7 & $+$18 24 08 & 17.7    & 16.8     & $ 0.74\pm0.03$* & 10695936 & \\
LP 726-11  & 8.68  & 8.01  & 7.63  & 7.53  & 08 48 36.4 & $-$13 53 08 & 17.5    & 17.5     & $ 1.07\pm0.22$  & 10700032 & \\
LP 491-51  & 8.49  & 8.03  & 7.55  & 7.47  & 11 03 21.2 & $+$13 37 57 & 17.7    & 16.4     & $ 0.74\pm0.23$  & 10701568 & \\
G 119-062  & 8.48  & 7.61  & 7.17  & 7.08  & 11 11 51.7 & $+$33 32 11 & 17.5    & 16.7     & $ 0.67\pm0.20$  & 10692864 & \\
LHS 2739   & 8.56  & 8.53  & 8.12  & 8.06  & 13 27 19.6 & $-$31 10 39 & 18.0    & 17.0     & $ 1.49\pm0.23$  & 10697472 & 1 \\
Gl 540.2   & 8.73  & 8.34  & 9.50  & 7.78  & 14 13 04.9 & $-$12 01 26 & 17.3    & 17.1     & $ 0.18\pm0.24$  & 10696192 & \\
Steph 1145 & 8.54  & 7.92  & 7.50  & 7.42  & 14 20 04.7 & $+$39 03 01 & 17.8    & 16.9     & $ 1.38\pm0.22$  & 10701824 & \\
Wo 9520    &\nodata&\nodata&\nodata&\nodata& 15 21 52.9 & $+$20 58 39 & \nodata & \nodata  & $ 0.28\pm0.05$* & \nodata  & \\
G 180-011  & 8.49  & 8.05  & 7.72  & 7.65  & 15 55 31.7 & $+$35 12 02 & 17.4    & 17.1     & $ 0.16\pm0.24$  & 10693888 & 1 \\
LP 331-57  & 7.05  & 7.02  & 7.03  & 6.88  & 17 03 52.7 & $+$32 11 45 & 17.7    & 16.9     & $ 0.57\pm0.34$  & 10698496 & 1,2 \\
LP  71-82  & 8.44  & 7.89  & 7.21  & 7.18  & 18 02 16.6 & $+$64 15 44 & 17.6    & 17.4     & $-0.73\pm0.19$  & 10699520 & \\ 
LP 390-16  & 8.53  & 8.17  & 7.73  & 7.67  & 18 13 06.6 & $+$26 01 52 & 17.9    & 17.3     & $ 0.61\pm0.25$  & 10699008 & \\
G 125-15   & 8.90  & 8.59  & 8.45  & 8.39  & 19 31 12.6 & $+$36 07 30 & 17.6    & 17.2     & $ 0.93\pm0.25$  & 10693120 & \\
LP 756-3   & 8.67  & 8.40  & 8.07  & 8.03  & 20 46 43.5 & $-$11 48 13 & 17.5    & 16.9     & $ 0.82\pm0.32$  & 10700288 & \\ 
BD-22 5866 & 6.65  & 6.86  & 6.61  & 6.52  & 22 14 38.4 & $-$21 41 53 & 17.8    & 17.0     & $ 1.58\pm0.34$  & 10692352 & 2 \\
Steph 2018 & 8.51  & 7.81  & 7.46  & 7.43  & 22 33 22.6 & $-$09 36 53 & 17.4    & 17.2     & $ 1.14\pm0.22$  & 10702336 & 1 \\
LP 984-91  & 6.69  & 6.69  & 6.61  & 6.60  & 22 44 57.9 & $-$33 15 01 & 17.5    & 16.7     & $ 1.87\pm0.18$  & 10701312 & 2 \\ 
Gl 873     & 5.59  & 5.11  & 5.00  & 4.92  & 22 46 49.7 & $+$44 20 03 & 18.0    & 17.3     & $-1.48\pm0.02$* & 10696448 & 2 \\
Gl 875.1   & 6.78  & 6.83  & 6.82  & 6.59  & 22 51 53.4 & $+$31 45 15 & 17.7    & 16.7     & $ 0.77\pm0.09$* & 10696704 & 2 \\
LP 642-48  & 8.73  & 8.33  & 8.11  & 8.18  & 23 20 57.5 & $-$01 47 37 & 17.6    & 17.3     & $ 1.23\pm0.30$  & 10699264 & 1 \\ 
LHS 4016   & 8.54  & 8.01  & 7.45  & 7.39  & 23 48 36.0 & $-$27 39 38 & 17.9    & 17.0     & $ 1.34\pm0.22$  & 10697728 & \\
G 68-46    & 8.86  & 8.61  & 8.47  & 8.46  & 23 51 22.2 & $+$23 44 20 & 17.9    & 17.3     & $ 1.29\pm0.40$  & 10695168 & \\
LP 764-40  & 8.41  & 7.83  & 7.37  & 7.29  & 23 58 13.6 & $-$17 24 33 & 17.6    & 17.0     & $ 1.35\pm0.30$  & 10700544 & 1 \\

\enddata
\tablecomments{1) Resolved binary system described in \citet{daem07} and references therein. 2) IRAC data taken in HDR mode. 3) Primary saturated in 4.5$\mu$m channel. 4) Primary saturated in 3.6$\mu$m and 4.5$\mu$m channels and mildly saturated in 5.8$\mu$m and 8.0$\mu$m channels. 5) Wide companion to LHS 5133.  *Distance moduli are derived from photometric and spectroscopic parallaxes save for those with asterisks which are from Hipparcos parallaxes.}
\label{tab:targs}
\end{deluxetable}

\begin{deluxetable}{cccccccccccc}
\tabletypesize{\footnotesize}
\rotate
\tablewidth{0pt}
\tablecaption{2MASS-IRAC Candidates}
\tablehead{
 \colhead{$m_{4.5}$} & \colhead{$3.6-4.5$} & \colhead{$4.5-5.8$} & \colhead{$5.8-8.0$} & \colhead{Distance} & \colhead{${m_{J}}^1$} & \colhead{${m_{H}}^1$} & \colhead{${m_{K}}^1$} & \colhead{RA} & \colhead{DEC} & \colhead{Primary} & \colhead{Follow Up Notes} \\
\colhead{(mag)} & \colhead{(mag)} & \colhead{(mag)} & \colhead{(mag)} & \colhead{(pc)} & \colhead{(mag)} & \colhead{(mag)} & \colhead{(mag)} & \colhead{(hms)} & \colhead{(dms)} & \colhead{} & \colhead{}
}

\startdata

 14.08 & 0.60 & \nodata & \nodata & 13.7 & 15.50 & 15.26 & 15.64 & 02 08 52.5 & +49 27 00.0 & G 173-39  & No CPM \\
  7.96$^2$ &\nodata$^2$& 0.29    & -0.01   & 15.6 &  8.90 &  8.37 &  8.06 & 07 28 51.1 & -30 15 53.8 & GJ 2060   & CPM Companion: SpT M5 \\
 14.01 & 0.08 &   -0.04 &   -0.01 & 18.0 & 15.06 & 14.56 & 14.47 & 23 58 05.6 & -17 23 39.6 & LP 764-40 & SpT M4: Not a Companion \\

\enddata
\label{tab:tmc}
\tablecomments{1) Magnitudes from the 2MASS database.}
\tablecomments{2) Saturated in Channel 1 and nearing saturation in Channel 2.}

\end{deluxetable}

\begin{deluxetable}{ccccccccccc}
\tabletypesize{\footnotesize}
\rotate
\tablewidth{0pt}
\tablecaption{IRAC-only Final Candidates}
\tablehead{
 \colhead{$m_{4.5}$} & \colhead{$3.6-4.5$} & \colhead{$4.5-5.8$} & \colhead{$5.8-8.0$} & \colhead{$M_{4.5}$} & \colhead{$m_{J}$} & \colhead{$m_{K}$} & \colhead{RA(hms)} & \colhead{DEC(dms)} & \colhead{Primary} & \colhead{Follow Up Notes} }

\startdata

 15.24 & 0.69 &    0.45 &    0.37 & 13.59 & 19.23   & 17.21   & 01 52 58.6 & -21 06 03.3 & LP 828-89 & PANIC: Wrong NIR Color  \\
 15.72 & 0.87 & \nodata & \nodata & 14.07 & 18.27   & 16.96   & 01 53 27.2 & -21 05 18.2 & LP 828-89 & PANIC: Source Elongated \\
 15.75 & 1.86 &    0.27 &    1.96 & 14.68 & 18.09   & 16.38   & 02 02 53.3 & +13 34 26.3 & G 003-035 & PANIC: Wrong NIR Color  \\
 14.86 & 2.62 & \nodata & \nodata & 13.79 & 20.30   & 18.55   & 02 02 59.4 & +13 33 34.9 & G 003-035 & PANIC: Source Elongated \\
 15.79 & 1.00 &    0.40 & \nodata & 14.24 & \nodata & \nodata & 03 49 46.2 & +24 21 38.5 & II Tau    & SWIRC: Source Elongated \\
 14.49 & 1.94 & \nodata & \nodata & 13.67 & 18.22   & 16.66   & 20 46 53.7 & -11 45 46.7 & LP 756-3  & PANIC: Wrong NIR Color  \\
 16.00 & 1.27 & \nodata & \nodata & 14.44 & 20.65   & 18.83   & 22 14 41.9 & -21 38 15.4 & BD-22 5866& PANIC: Wrong NIR Color  \\
 15.64 & 2.70 &    0.49 & \nodata & 13.77 & 18.29   & 16.60   & 22 45 03.8 & -33 16 29.0 & LP 984-91 & PANIC: Wrong NIR Color  \\
 15.55 & 2.60 &    0.54 &    2.43 & 14.79 & \nodata & \nodata & 22 52 02.4 & +31 44 11.5 & Gl 875.1  & SWIRC: Source Elongated \\
 14.81 & 1.00 &   -0.10 & \nodata & 13.52 & 15.78   &14.51$^1$& 23 51 26.0 & +23 42 44.3 & G 68-46   & Spex:  Wrong NIR Colors \\

\enddata
\label{tab:final}
\tablecomments{1) H band magnitude not K}

\end{deluxetable}

\begin{deluxetable}{ccc}
\tablewidth{0pt}
\tablecaption{GJ 2060 system}
\tablehead{
 \colhead{Quantity} & \colhead{GJ 2060AB} & \colhead{GJ 2060C} }

\startdata

m$_J$           & 6.62 mag                 & 8.90 mag \\
m$_H$           & 5.97 mag                 & 8.37 mag \\
m$_Ks$          & 5.72 mag                 & 8.06 mag \\
PM$_{\alpha}^1$ & -105~$\frac{mas}{yr}$ & -131~$\frac{mas}{yr}$ \\
PM$_{\delta}^1$ & -209~$\frac{mas}{yr}$ & -209~$\frac{mas}{yr}$ \\
Distance        & 15.6~pc                  & \nodata \\
SpT             & M1V                      & M5V$^2$ \\

\enddata
\label{tab:gj}
\tablecomments{1) Proper motions were measured for both objects using 2MASS as a first epoch and our IRAC data as a second epoch. \\ 2) Based on a low resolution Spex spectrum.}
\end{deluxetable}

\end{document}